\pdfoutput=1
\documentclass[runningheads]{llncs}
\usepackage{graphicx}
\usepackage{multirow}
\usepackage{array}
\usepackage{amsmath}
\usepackage{amssymb}
\usepackage[strings]{underscore}
\usepackage[misc]{ifsym}
\usepackage[colorlinks]{hyperref}

\begin{document}
\title{TarGAN: Target-Aware Generative Adversarial Networks for Multi-modality Medical\\Image Translation}
\titlerunning{TarGAN}

\author{
Junxiao Chen\inst{1} \and
Jia Wei\inst{1}\textsuperscript{(\Letter)} \and
Rui Li\inst{2}
}
\authorrunning{J. Chen et al.}
\institute{
\textsuperscript{1}School of Computer Science and Engineering, South China University of Technology, Guangzhou, China\\
\email{cs\_xiao@mail.scut.edu.cn},
\email{csjwei@scut.edu.cn}\\
\textsuperscript{2}Golisano College of Computing and Information Sciences, Rochester Institute of Technology, Rochester, NY 14623\\
\email{rxlics@rit.edu}
}
\maketitle              
\begin{abstract}
Paired multi-modality medical images, can provide complementary information to help physicians make more reasonable decisions than single modality medical images. But they are difficult to generate due to multiple factors in practice (e.g., time, cost, radiation dose). To address these problems, multi-modality medical image translation has aroused increasing research interest recently. However, the existing works mainly focus on translation effect of a whole image instead of a critical target area or Region of Interest (ROI), e.g., organ and so on. This leads to poor-quality translation of the localized target area which becomes blurry, deformed or even with extra unreasonable textures. 
In this paper, we propose a novel target-aware generative adversarial network called \textbf{TarGAN}, which is a generic multi-modality medical image translation model capable of (1) learning multi-modality medical image translation without relying on paired data, (2) enhancing quality of target area generation with the help of target area labels. The generator of TarGAN jointly learns mapping at two levels simultaneously — whole image translation mapping and target area translation mapping. These two mappings are interrelated through a proposed crossing loss. The experiments on both quantitative measures and qualitative evaluations demonstrate that TarGAN outperforms the state-of-the-art methods in all cases. Subsequent segmentation task is conducted to demonstrate effectiveness of synthetic images generated by TarGAN in a real-world application. Our code is available at \url{https://github.com/2165998/TarGAN}.
\keywords{Multi-modality translation \and GAN \and Abdominal organs.}
\end{abstract}
\section{Introduction}
Medical imaging, a powerful diagnostic and research tool creating visual representations of anatomy, has been widely available for disease diagnosis and surgery planning \cite{ernst2019cnn}. In current clinical practice, Computed Tomography (CT) and Magnetic Resonance Imaging (MRI) are most commonly used.
Since CT and multiple MR imaging modalities provide complementary information, an effective integration of these different modalities can help physicians make more informative decisions. 

Since it is difficult and costly to obtain paired multi-modality images in clinical practice, there is a growing demand for developing multi-modality image translations to assist clinical diagnosis and treatment \cite{xin2020multi}.

\begin{figure}
\centering
\includegraphics[width=\textwidth]{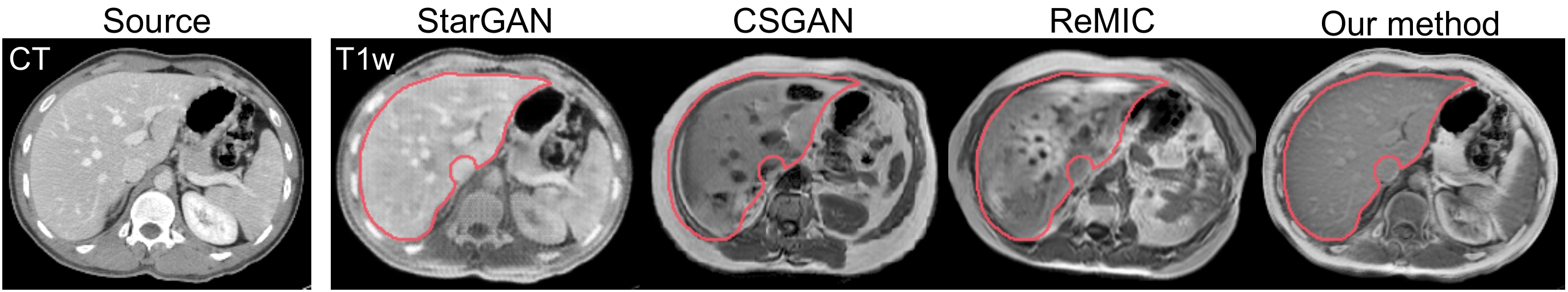}
\caption{Translation results (CT to T1w) of different methods are shown here. The target area (i.e., liver) is contoured in red.}
\label{fig:5}
\end{figure}

Existing works can be categorized into two types. One is crossing-modality medical image translation between two modalities, which has scalability issues to the increasing number of modalities \cite{yu2019ea,zhang2018translating}, since these methods have to train $n(n-1)$ generator  models in order to learn all mappings between $n$ modalities. The other is multi-modality image translation \cite{choi2018stargan,huang2019coca,shen2020multi,xin2020multi}. In this category, some methods \cite{huang2019coca,xin2020multi} rely on paired data, which is hard to acquire in clinical reality. Other methods \cite{choi2018stargan,shen2020multi} can learn from unpaired data, however, they tend to lead to deformation in target area without prior knowledge, as concluded by Zhang \emph{et al.} \cite{zhang2018translating}. 
As demonstrated in Figure \ref{fig:5}, the state-of-the-art multi-modality image translation methods give rise to poor quality local translations. The translated target area (For example, Liver, in red curves) is blurry, deformed or perturbed with redundant unreasonable textures. Comparing to them, our method can not only perform whole image translation in competitive quality but also achieve significantly better local translation for the target area.

To address the above issues, we present a novel unified general-purpose multi-modality medical image translation method named ``Target-Aware Generative Adversarial Networks'' (TarGAN). We incorporate target labels to enable the generator to focus on local translation of target area. The generator has two input-output streams. One stream translates a whole image from source modality to target modality, the other focuses on translating a target area. In particular, we combine the cycle-consistency loss \cite{zhu2017unpaired} and the backbone of StarGAN \cite{choi2018stargan} to learn the generator, which enables our model to scale up to modality increase without relying on paired data. Then, the untraceable constraint \cite{zhu2019ugan} is employed to further improve translation quality of synthetic images. To avoid the deformation of output images caused by untraceable constraint, we construct a shape-consistency loss \cite{fu2018three} with 
an auxiliary network, namely shape controller. We further propose a novel crossing loss to allow the generator to focus on the target area when translating the whole image to target modality. Trained in an end-to-end fashion, TarGAN can not only accomplish multi-modality translation but also properly retain the target area information in the synthetic images. 

\subsubsection{Overall, the contributions of this work are:}
(1) We propose TarGAN to generate multi-modality medical images with high-quality local translation on target areas by integrating global and local mappings with a crossing loss.
(2) We show qualitative and quantitative performance evaluations on multi-modality medical image translation tasks with CHAOS2019 dataset \cite{kavur2019chaos}, demonstrating our method's superiority over the state-of-the-art methods.
(3) We further use the synthetic images generated from TarGAN to improve the performance of a segmentation task, which indicates that the synthetic images generated by TarGAN achieve the improvement by enriching the information of source images.
\begin{figure}
\includegraphics[width=\textwidth]{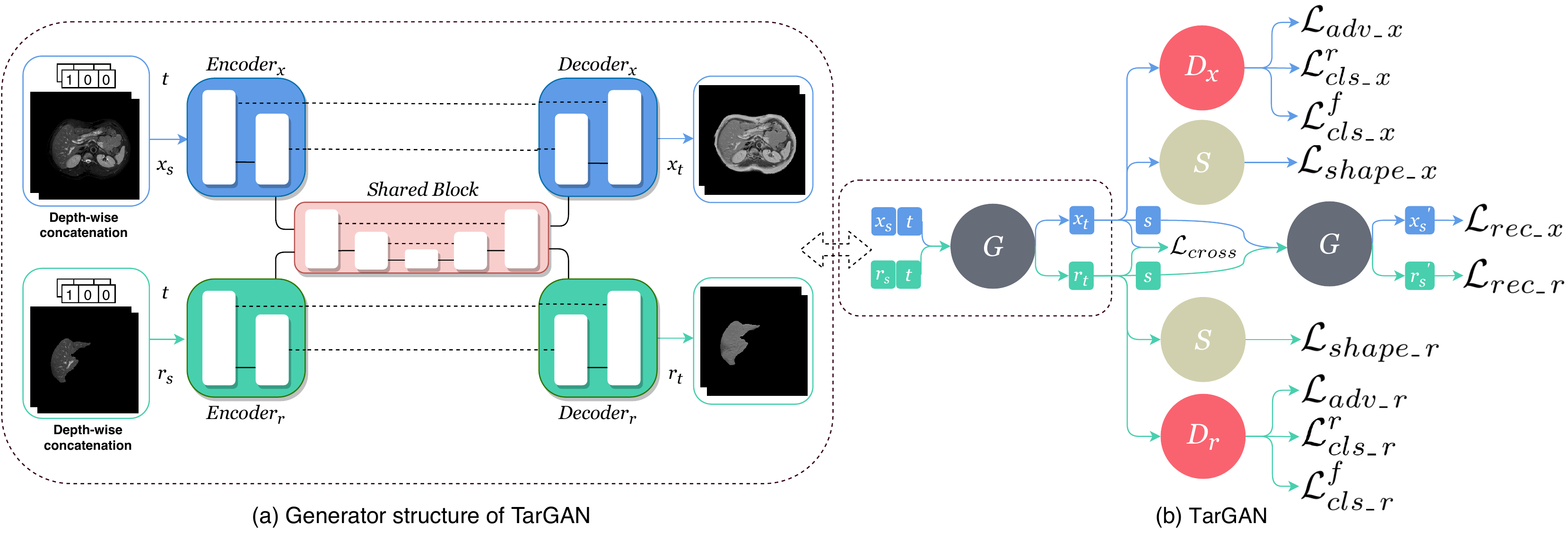}
\caption{The illustration of TarGAN. As in \textbf{(b)}, TarGAN consists of four modules ($G$, $S$, $D_x$, $D_r$). The generator $G$ translates a source whole image $x_s$ and a source target area image $r_s$ to a target whole image $x_t$ and a target area image $r_t$. The detailed structure of $G$ is shown in  \textbf{(a)}. The shape controller $S$ preserves the invariance of anatomy structures. The discriminators $D_x$ and $D_r$ distinguish whether a whole image and its target area are real or fake and determine which modalities the source images come from.}
\label{fig:gan overview}
\end{figure}
\section{Methods}
\label{se:to}
\subsection{Proposed framework}
Given an image $x_{s}$ from source modality $s$ and its corresponding target area label $y$, we specify a target area image $r_s$ which only contains the target area by binarization operation $y \cdot x_{s}$. Given any target modality $t$, our goal is to train a single generator $G$ that can translate any input image $x_s$ of source modality $s$ to the corresponding output image $x_t$ of target modality $t$, and translate the input target area image $r_s$ of source modality $s$ to the corresponding output target area image $r_t$ of target modality $t$ simultaneously, denoted as $G(x_s, r_s, t) \rightarrow (x_t, r_t)$. Figure \ref{fig:gan overview} illustrates the architecture of TarGAN, which is composed of four modules described below.

To achieve the aforementioned goal, we design a double input-output streams \textbf{generator $G$} consisting of a shared middle block and two pairs of encoder-decoder. Combining with the shared middle block, both encoder-decoder pairs translate an input image into an output image of the target modality $t$. One stream's input is the whole image $x_{s}$, and the other's input only includes the target area $r_s$. The shared middle block is designed to implicitly enable $G$ to focus on target area in whole image translation.
Note that target area label $y$ of $x_s$ is not available in test phase, so the input block $Encoder_r$ and output block $Decoder_r$ are not used at that time.

Given a synthetic image $x_t$ or $r_{t}$ from ${G}$, \textbf{the shape controller $S$} generates a binary mask which can represent the foreground area of the synthetic image.
 
Lastly, we use \textbf{two discriminators denoted as {$D_{x}$} and {$D_{r}$}} corresponding to two output streams of {$G$}. The probability distributions inferred by $D_{x}$ distinguish whether the whole image is real or fake, and determine which modality the whole image comes from. Similarly, the $D_{r}$ distinguish whether the target area image is real or fake, and to determine which modality the target area image comes from.
\subsection{Training objectives}
\subsubsection{Adversarial loss.} To minimize the difference between the distributions of generated images and real images, we define the adversarial loss as
\begin{equation}
\begin{split}
\mathcal{L}_{adv\_\,x} = &\,\,\mathbb{E}_{x_s}[\textrm{log}\,\emph{D}_{src\_\,x}(x_s)]\,+\,\mathbb{E}_{x_t}[\textrm{log}(1\,-\,\emph{D}_{src\_\,x}(x_t))],
\\
\mathcal{L}_{adv\_\,r} = &\,\,\mathbb{E}_{r_s}[\textrm{log}\,\emph{D}_{src\_\,r}(r_s)]\,+\,\mathbb{E}_{r_t}[\textrm{log}(1\,-\,\emph{D}_{src\_\,r}(r_t))].
\end{split}
\end{equation}
Here, $D_{src\_\,x}$ and $D_{src\_\,r}$ represent the probability distributions of real or fake over input whole images and target area images.
\subsubsection{Modality classification loss.} To assign the generated image to their target modality $t$, we impose the modality classification loss on $G$, $D_{x}$ and $D_{r}$. The loss consists of two terms: modality classification loss of
real images which is used to optimize $D_x$ and $D_r$, denoted as $\mathcal{L}_{cls\_(x/r)}^r$, and modality classification loss of fake images which is used to optimize $G$, denoted as $\mathcal{L}_{cls\_(x/r)}^f$.
In addition, to eliminate synthetic images' style features from source modalities,
the untraceable constraint \cite{zhu2019ugan} is combined into $\mathcal{L}_{cls\_(x/r)}^r$ as:
\begin{equation}
\begin{split}\label{loss:dc}
\mathcal{L}_{cls\_\,x}^{r} = &\,\,\mathbb{E}_{x_{s},s}[-\textrm{log}\,\emph{D}_{cls\_\,x}(s|x_{s})]\,+\lambda_{u}\,\mathbb{E}_{x_t,s'}[-\textrm{log}\,\emph{D}_{cls\_\,x}(s'|x_t)],
\\
\mathcal{L}_{cls\_\,r}^{r} = &\,\,\mathbb{E}_{r_s,s}[-\textrm{log}\,\emph{D}_{cls\_\,r}(s|r_s)]\,+\lambda_{u}\,\mathbb{E}_{r_t,s'}[-\textrm{log}\,\emph{D}_{cls\_\,r}(s'|r_t)].
\end{split}
\end{equation}
Here, $D_{cls\_\,x}$ and $D_{cls\_\,r}$ represent the probability distributions over modality labels and input images. $s'$ indicates whether an input image is fake, and is translated from a source modality $s$ \cite{zhu2019ugan}. Besides, we define $\mathcal{L}_{cls\_(x/r)}^{f}$ as
\begin{equation}
\mathcal{L}_{cls\_\,x}^{f} = \mathbb{E}_{x_t,t}[-\textrm{log}\,\emph{D}_{cls\_\,x}(t|x_t)],\,\,
\mathcal{L}_{cls\_\,r}^{f} = \mathbb{E}_{r_{t},t}[-\textrm{log}\,\emph{D}_{cls\_\,r}(t|r_{t})].
\end{equation}
\subsubsection{Shape consistency loss.}
Since the untraceable constraint can affect the shape of anatomy structures in synthetic images by causing structure deformation, 
we correct it by adding a shape consistency loss \cite{fu2018three} to $G$ with shape controller $S$ as
\begin{equation}
\mathcal{L}_{shape\_\,x} = \mathbb{E}_{x_t, b^x}[||b^x - S(x_t)||_{2}^{2}],
\,\,
\mathcal{L}_{shape\_\,r} = \mathbb{E}_{r_t, b^r}[||b^r - S(r_t)||_{2}^{2}],
\end{equation}
where $b^x$ and $b^r$ are the binarizations (with 1 indicating foreground pixels and 0 otherwise) of $x_s$ and $r_s$. $S$ constrains $G$ to focus on the multi-modality mapping in a content area.
\subsubsection{Reconstruction loss.} To allow $G$ to preserve the modality-invariant characteristics of the whole image $x_s$ and its target area image $r_s$, we employ a cycle consistency loss \cite{zhu2017unpaired} as
\begin{equation}
\mathcal{L}_{rec\_\,x} = \mathbb{E}_{x_{s}, x_s'}[||x_s - x_s'||_{1}],
\,\,
\mathcal{L}_{rec\_\,r} = \mathbb{E}_{r_s, r_s'}[||r_s - r_s'||_{1}].
\end{equation}
Note that $x_s'$ and $r_s'$ are from $G(x_t,r_t,s)$. Given the paired synthetic image ($x_t,  r_t$) and the source modality $s$, $G$ tries to reconstruct the input images ($x_s, r_s$).
\subsubsection{Crossing loss.} To enforce $G$ to focus on a target area when generating a whole image $x_t$, we directly regularize $G$ with a crossing loss defined as
\begin{align}
\mathcal{L}_{cross} = \mathbb{E}_{x_t, r_t, y}[||x_t \cdot y - r_t||_1],
\label{loss:crossing}
\end{align}
where $y$ is the target area label corresponding to $x_s$. By minimizing the crossing loss, $G$ can jointly learn from double input-output streams and share information between them.
\subsubsection{Complete objective.} By combining the proposed losses together, our complete objective functions are as follows:
\begin{align}
\mathcal{L}_{D_{(x/r)}} = - \mathcal{L}_{adv\_(x/r)} + \lambda_{cls}^r\,\mathcal{L}_{cls\_(x/r)}^r,
\end{align}
\begin{equation}
\begin{aligned}
\mathcal{L}_{G} &= \mathcal{L}_{adv\_(x/r)} + \lambda_{cls}^f\,\mathcal{L}_{cls\_(x/r)}^f + \lambda_{rec}\,\mathcal{L}_{rec\_(x/r)} + \lambda_{cross}\,\mathcal{L}_{crossing},
\end{aligned}
\end{equation}
\begin{align}
\mathcal{L}_{G,S} = \mathcal{L}_{shape\_(x/r)},
\end{align}
where $\lambda_{cls}^r$, $\lambda_{cls}^f$, $\lambda_{rec}$, $\lambda_{cross}$ and $\lambda_u$ (Eqs. (\ref{loss:dc})) are hyperparameters to control the relative importance of each loss. 
\begin{table}
\caption{Quantitative evaluations on synthetic images of different methods. ($\uparrow$ denotes higher is better, while $\downarrow$ denotes lower is better)}
\centering
\begin{tabular}{l|c|c|c|c|c|c}
\hline
\multirow{2}{*}{Method} & \multicolumn{3}{c|}{FID$\downarrow$} & \multicolumn{3}{c}{S-score(\%)$\uparrow$}\\
\cline{2-7} & CT & T1w & T2w & CT & T1w & T2w\\
\hline
StarGAN \cite{choi2018stargan}  &0.0488&0.1179&0.2615&42.89&29.23&42.17\\
CSGAN \cite{zhang2018translating} 	 &0.0484&0.1396&0.4819&56.72&45.67&69.09\\
ReMIC \cite{shen2020multi} 		&0.0912&0.1151&0.5925&51.03&32.00&69.58\\
Our method                 		&\textbf{0.0418}&\textbf{0.0985}&\textbf{0.2431}&\textbf{57.13}&\textbf{65.79}&\textbf{69.63}\\
\hline
\end{tabular}
\label{fs}
\end{table}
\section{Experiments and results}
\subsection{Settings}
\subsubsection{Dataset.} We use 20 patients' data in each modality (CT, T1-weighted and T2-weighted). They are from the Combined Healthy Abdominal Organ Segmentation (CHAOS) Challenge \cite{kavur2021chaos}. Detailed imaging parameters are shown in supplementary material.
We resize all slices as $256 \times 256$ uniformly. $50\%$ data from each modality are randomly selected as training data, while the rest as test data. Because CT scans only have liver labels, we set liver as the target area.
\subsubsection{Baseline methods.}
Translation results comparisons are conducted against the state-of-the-art
translation methods, StarGAN \cite{choi2018stargan}, CSGAN \cite{zhang2018translating} and ReMIC \cite{shen2020multi}. Note that we implement an unsupervised ReMIC because of the lack of ground-truth images.

Target segmentation performances are also evaluated against the above methods. We train and test models using only real images of each modality, denoted as \textbf{Single}. We use the mean results of two segmentation models of each modality from CSGAN and use the segmentation model $G_s$ from ReMIC. As for StarGAN and TarGAN, inspired by \emph{`image enrichment'} \cite{gupta2019gan}, we extend every single modality to multiple modalities and concatenate multiple modalities within each sample, as [CT] $\rightarrow$ [CT, synthetic T1w, synthetic T2w].
\subsubsection{Evaluation metrics.}
In the translation tasks, due to the lack of ground-truth images, we can not use the common metrics like PSNR, SSIM, etc. So we evaluate both the visual quality and the integrity of target area structures of generated images using Frechét inception distance (\textbf{FID}) \cite{heusel2017gans} and segmentation score
(\textbf{S-score}) \cite{zhang2018translating}. We compute FID and S-score for each modality and report their average values. The details on above metrics are further described in supplementary material.

In the segmentation tasks, dice coefficient (\textbf{DICE}) and relative absolute volume difference (\textbf{RAVD}) are used as metrics. We compute each metric on every modality, and report their average values and standard deviations.
\label{sec:id}
\subsubsection{Implementation details.}We use U-net \cite{ronneberger2015u} as the backbone of $G$ and $S$. In $G$, only half of the channels are used for every skip connection. As for $D_x$ and $D_r$, we implement the backbone with PatchGAN \cite{isola2017image}. Details of above networks are included in the supplementary material. All the liver segmentation experiments are conducted with nnU-Net \cite{isensee2021nnu} except CSGAN and ReMIC.

To stabilize the training process, we adopt Wasserstein GAN loss with a gradient penalty \cite{gulrajani2017improved,martin2017wasserstein} using 
$\lambda_{gp} = 10$ and two-timescale update rule (TTUR) \cite{heusel2017gans} for $G$ and $D$. The learning rates for $G$, $S$ are
set to $10^{-4}$, while that of $D$ is set to $3 \times 10^{-4}$. We set $\lambda_{cls}^r=1$, $\lambda_{cls}^f=1$, $\lambda_{rec} = 1$, $\lambda_{cross} = 50$ and $\lambda_u = 0.01$. The batch size and training epoch are set to $4$ and $50$, respectively. We use the Adam optimizer \cite{kingma2015adam} with momentum parameters $\beta_1 = 0.5$ and $\beta_2 = 0.9$. All images are normalized to [$-1, 1$] prior to the training and test. We use exponential moving averages over parameters \cite{karras2018progressive} of $G$ during test, with a decay of $0.999$. Our implementation is trained on an NVIDIA GTX 2080Ti with PyTorch.
\begin{figure}
\centering
\includegraphics[width=\textwidth]{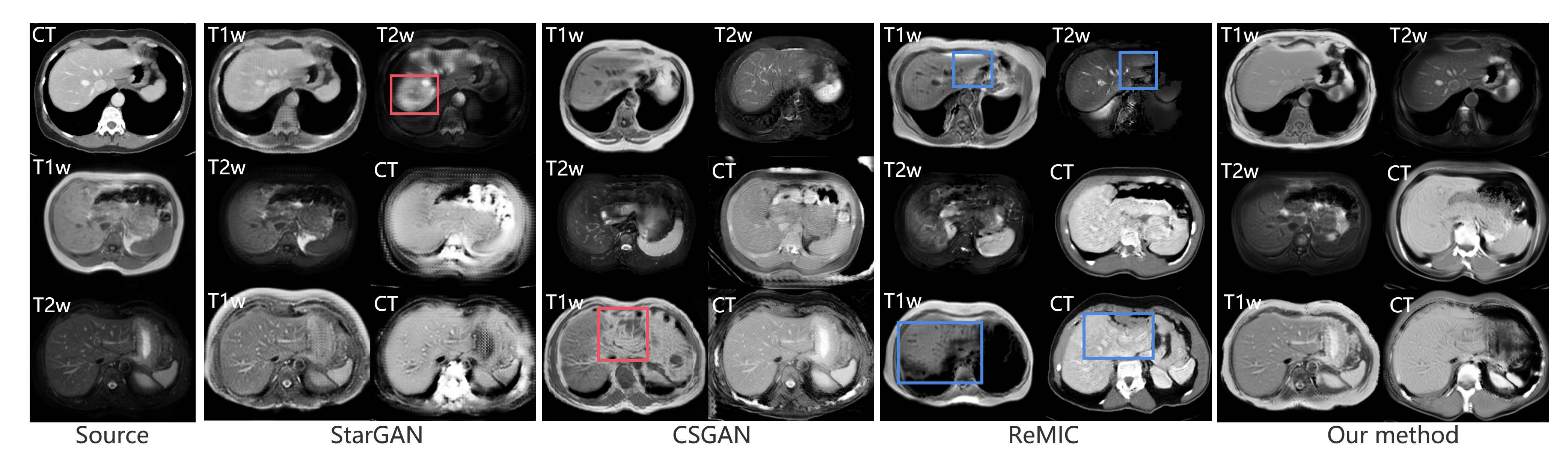}
\caption{Multi-modality medical image translation results. Red boxes highlight the redundant textures, and blue boxes indicate the deformed structures.}
\label{qualitativeImgs}
\end{figure}
\subsection{Results and analyses}
\subsubsection{Image Translation.} Figure \ref{qualitativeImgs} shows qualitative results on each pair of modal image translation. As shown, StarGAN fails to translate image from CT to T1w and produces many artifacts in MRI to CT translation. CSGAN sometimes adds redundant textures (marked by the red boxes) in the target area while retaining the shape of target. ReMIC tends to generate relatively 
realistic synthetic images while deforming the structure of target area in most cases (marked by the blue boxes).
Comparing to above methods, TarGAN generates translation results in higher visual quality and properly preserves the target structures. Facilitated by the proposed crossing loss, TarGAN can jointly learn the mappings of the target area and the whole image among different modalities, and further make $G$ focus on the target areas to improve their quality. Furthermore, as shown in Table \ref{fs}, TarGAN outperforms all the baselines in terms of FID and S-score, which suggests TarGAN produces the most realistic medical images, and the target area integrity of synthetic images derived from TarGAN is significantly better.
\begin{table}[!h]
\caption{Liver segmentation results (mean $\pm$ standard deviation) on different medical modalities.}
\label{seg}
\centering
\begin{tabular}{l|c|c|c|c|c|c}
\hline
\multirow{2}{*}{Method}
&\multicolumn{3}{c|}{DICE($\%$)$\uparrow$} 
&\multicolumn{3}{c}{RAVD($\%$)$\downarrow$}\\
\cline{2-7}
& CT & T1w& T2w
& CT & T1w& T2w\\
\hline
Single &96.29$\pm$0.74 &93.53$\pm$2.43 & 89.24$\pm$8.18 & 3.31$\pm$1.80 & 3.81$\pm$3.49& 11.68$\pm$14.37\\
StarGAN \cite{choi2018stargan}&96.65$\pm$0.34& 92.71$\pm$1.66& 86.38$\pm$4.95 &3.07$\pm$1.53&5.40$\pm$2.87& 15.71$\pm$9.85\\
CSGAN \cite{zhang2018translating}&96.08$\pm$2.05& 87.47$\pm$5.97& 86.35$\pm$6.29& 4.47$\pm$3.94& 15.74$\pm$14.18& \textbf{8.23$\pm$8.55}\\
ReMIC \cite{shen2020multi}& 93.81$\pm$1.43& 86.33$\pm$8.50& 82.70$\pm$4.36& 5.33$\pm$3.55 & 8.06$\pm$8.80& 10.62$\pm$6.80\\
Our method & \textbf{97.06$\pm$0.62}& \textbf{94.02$\pm$2.00}& \textbf{90.94$\pm$6.28} & \textbf{2.33$\pm$1.60}& \textbf{3.50$\pm$1.82}& 9.92$\pm$11.17\\
\hline
\end{tabular}
\end{table}
\subsubsection{Liver segmentation.}
The quantitative segmentation results are shown in Table \ref{seg}. Our method achieves better performance than all other methods on most of the metrics. This suggests TarGAN can not only generate realistic images for every modality, but also properly retain liver structure in synthetic images. The high-quality local translation for the target areas plays a key
role in the improvement of liver segmentation performance. By jointly learning from real and synthetic images, the segmentation models can incorporate more information on the liver areas within each sample.
\subsubsection{Ablation test.}
We conduct an ablation test to validate effectiveness of different parts of TarGAN in terms of preserving target area information. For ease of presentation, we denote \textbf{shape controller}, \textbf{target area translation mapping} and \textbf{crossing loss} as \textbf{S}, \textbf{T} and \textbf{C}, respectively. As shown in Table \ref{ss}, \textbf{TarGAN without (w/o) S, T, C} is closely similar to StarGAN except using our implementation. The proposed crossing loss plays a key role in TarGAN, which increases the mean of S-score from \textbf{TarGAN w/o C} $51.03\%$ to $64.18\%$.
\begin{table}
\caption{Ablation study on different components of TarGAN. Note that \textbf{TarGAN w/o S, T} and \textbf{TarGAN w/o T} don't exist, since \textbf{T} is the premise of \textbf{C}.}
\begin{center}
\begin{tabular}{l|c|c|c|c}
\hline
\multirow{2}{*}{Method} & \multicolumn{4}{c}{S-score($\%$)} \\ 
\cline{2-5}
&CT & T1w & T2w & Mean\\
\hline
TarGAN w/o S,\,T,\,C   &30.64&35.05&67.45&44.38\\
TarGAN w/o S,\,C    &39.78&29.96&67.47&45.74\\
TarGAN w/o T,\,C    &37.42&38.33&68.85&48.20\\
TarGAN w/o C      &43.00 &38.83 &71.27&51.03 \\
TarGAN w/o S      &56.69 &59.37 &\textbf{71.89}&62.65 \\
TarGAN         &\textbf{57.13}&\textbf{65.79}&69.63 &\textbf{64.18}\\
\hline
\end{tabular}
\end{center}
\label{ss}
\end{table}
\section{Conclusion}
In this paper, we propose a novel general-purpose method TarGAN to mainly address two challenges in multi-modality medical image translation: learning multi-modality medical image translation without relying on paired data, and improving the quality of local translation on target area. A novel translation mapping mechanism is introduced to enhance the target area quality during generating the whole image. Additionally, by using the shape controller to alleviate the deformation problem caused by the untraceable constraint and combining a novel crossing loss in generator $G$, TarGAN addresses both challenges within a unified framework. Both the quantitative and qualitative evaluations show the superiority of TarGAN in comparison with the state-of-the-art methods. We further conduct a segmentation task to demonstrate effectiveness of synthetic images generated by TarGAN in a real application.
\subsubsection{Acknowledgments.}This work is supported in part by the Natural Science Foundation of Guangdong Province (2017A030313358, 2017A030313355, 2020A1515010717), the Guangzhou Science and Technology Planning Project (201704030051), the Fundamental Research Funds for the Central Universities (2019MS073), NSF-1850492 (to R.L.) and NSF-2045804 (to R.L.).
%
%
 \bibliographystyle{splncs04}
 \bibliography{targan}

\end{document}